\documentclass[aps,preprint]{revtex4}%
\usepackage{amsfonts}
\usepackage{amsmath}
\usepackage{amssymb}
\usepackage[final,truetex]{graphicx}%
\begin{document}
\preprint{hep-th/0404059}
\title[Short title for running header]{Mass singularity and confining property in QED$_{3}$ }
\author{Yuichi Hoshino}
\affiliation{Kushiro National College of Technology,Otanoshike nishi 2-32-1,Kushiro ,Hokkaodo,084-0096,Japan}

\begin{abstract}
Shell document for REV\TeX{} 4.

\end{abstract}
\begin{abstract}
We discuss the properties of the position space fermion propagator in three
dimensional QED which has been found previously based on
Ward-Takahashi-identity for soft-photon emission vertex and spectral
representation in quenched approximation.There is a new type of mass
singularity which governs the long distance behaviour.It leads the propagator
vanish at large distance.This term corresponds to dynamical mass in position
space and expressed as superposition of the propagator with different mass in
momentum space.Our model shows confining property and dynamical mass
generation for arbitrary coupling constant.Since we used dispersion relation
in deriving spectral function there is a physical mass which sets a mass
scale. Low energy behaviour of the proagator is modified to decrease by
position dependent mass.In the limit of zero infrared cut-off the propagator
vanishes with new kind of infrared behaviour.

\end{abstract}
\volumeyear{year}
\volumenumber{number}
\issuenumber{number}
\eid{identifier}
\date[Date text]{date}
\received[Received text]{date}

\revised[Revised text]{date}

\accepted[Accepted text]{date}

\published[Published text]{date}

\startpage{101}
\endpage{121}
\maketitle
\tableofcontents

\section{Introduction}

Infrared behaviour of the propagator in the presence of massless particle
(i.e.photon and graviton) had been discussed assuming spectral representation
of the propagator and the use of Ward-Takahashi-identity[1].It is pointed out
that the spectral representation is model independent and the asymptotic form
of the scalar propagator in QED and Gravity theory were determined[1].It has
long been thought that the three dimensional QED is a confining theory since
its infrared divergences is severe[3].However it has not been known what we
can observe as a confinement.One of these examples is a vanishment of the
propagator in the infrared region.In the previous paper we find that the
scalar and spinor propagator have a new type of mass singularity in the same
approach in ref[1] to QED$_{3}$[2]. There are two kinds of gauge invariant
mass singularities in the evaluation of the $O(e^{2})$ spectral function by
LSZ reduction formula.After exponentiation of $O(e^{2})$ spectral function we
get the full propagator in position space which is a product of free
propagator with physical mass and quantum correction.In our model quantum
corrections are Coulomb energy and position dependent mass which are both
logarithmically divergent at long distance.Here confinement means that the
propagator damps faster than the free propagator with physical mass for
arbitrary coupling constant.First we study the structure of the propagator
with finite infrared cut-off $\mu.$It is assumed that the three dimensional
analysis leads the leading order of high temperature expansion results in four
dimension[3].Using Laplace transformation of the position dependent mass
numerically,we have the integral reresentation of the propagator in momentum
space as in dispersion theory.As a result we find in momentum space that the
propagator vanishes in the limit of zero infrared cut-off $\mu^{2}=m^{2}%
-p^{2}$ near $p^{2}=m^{2}.$In section II Bloch-Nordsieck approximation in
three dimension is reviewed.In section III mass singularity in four and three
dimension is compared and numerical analysis in momentum space is
given.Section IV is for non-perturbative effects as renormalization
constant,bare mass which are defined in the high-energy limit,and the vacuum
expectation value of the composite operator $\left\langle \overline{\psi}%
\psi\right\rangle .$Renormalization constant $Z_{2}^{-1}$ vanishes for
arbitrary coupling constant.There seems to be a critical coupling constant
which separates the phases of $\left\langle \overline{\psi}\psi\right\rangle
.$

\section{Ward-Takahashi-identity and the spectral function}

\subsection{Position space propagator}

First we consider about the charged particles which emitt and absorb massless
photons.Usually this process was described by spectral function;transition
probability of particle into particle and photon state.This method is model
independent and helpful in any dimension and leads to the so called
Bloch-Nordsieck approximation.Multi-photon emitted from external line is
introduced by ladder type diagrams which satisfy Ward-Takahashi-identity.Let
us begin by dispersion theoretic description of the propagator[1]
\begin{align}
S_{F}(p) &  =\int d^{3}x\exp(-ip\cdot x)\left\langle \Omega|T\psi
(x)\overline{\psi}(0)|\Omega\right\rangle \nonumber\\
&  =\int d\omega^{2}\frac{\gamma\cdot p\rho_{1}(\omega^{2})+\rho_{2}%
(\omega^{2})}{p^{2}-\omega^{2}+i\epsilon},
\end{align}%
\begin{align}
\frac{1}{\pi}\Im S(p) &  =\int d\omega^{2}\delta(p^{2}-\omega^{2})[\gamma\cdot
p\rho_{1}(\omega^{2})+\rho_{2}(\omega^{2})]\nonumber\\
&  =\gamma\cdot p\rho_{1}(p)+\rho_{2}(p),
\end{align}%
\begin{equation}
S_{F}(x)=\left\langle \Omega|T\psi(x)\overline{\psi}(0)|\Omega\right\rangle .
\end{equation}
The field $\psi$ is renormalized and is taken to be a spinor with mass
$m$.Here we introduce intermediate states that contribute to the spectral
function
\begin{align}
\rho(p) &  =\gamma\cdot p\rho_{1}(p)+\rho_{2}(p)\nonumber\\
&  =(2\pi)^{2}\sum_{N}\delta^{3}(p-p_{N})\int d^{3}x\exp(-ip\cdot
x)\left\langle \Omega|\psi(x)|N\right\rangle \left\langle N|\overline{\psi
}(0)|\Omega\right\rangle .
\end{align}
Total three-momentum of the state $|N\rangle$ is $p_{N}^{\mu}.$The only
intermediates $N$ contain one spinor and an arbitrary number of
photons.Setting
\begin{equation}
|N\rangle=|r;k_{1},...,k_{n}\rangle,
\end{equation}
where $r$ is the momentum of the spinor $r^{2}=m^{2},$and $k_{i}$ is the
momentum of $i$th soft photon,we have
\begin{align}
\rho(p^{2}) &  =\int\frac{md^{2}r}{r^{0}}\sum_{n=0}^{\infty}\frac{1}{n!}%
\times(\int\frac{d^{3}k}{(2\pi)^{3}}\theta(k_{0})\delta(k^{2})\sum_{\epsilon
})_{n}\delta(p-r-\sum_{i=1}^{n}k_{i})\nonumber\\
&  \times\left\langle \Omega|\psi(x)|r;k_{1},..,k_{n}\right\rangle
\left\langle r;k_{1},..k_{n}|\overline{\psi}(0)|\Omega\right\rangle .
\end{align}
Here the notations
\begin{equation}
(f(k))_{0}=1,(f(k))_{n}=\prod_{i=1}^{n}f(k_{i})
\end{equation}
have been introduced to denote the phase space integral of each photon.Intial
sum over $\epsilon$ is a sum over polarization of photon.To evaluate the
contribution of soft-photons,we consider when only $n$th photon is soft.Here
we define following matrix element
\begin{equation}
T_{n}=\left\langle \Omega|\psi|r;k_{1},..k_{n}\right\rangle .
\end{equation}
We consider $T_{n}$ for $k_{n}^{2}\neq0,$we continue off the photon mass-shell
by Lehmann-Symanzik-Zimmermann(LSZ)formula:%
\begin{equation}
T_{n}=\epsilon_{n}^{\mu}T_{n\mu},
\end{equation}%
\begin{align}
\epsilon_{n}^{\mu}T_{n}^{\mu} &  =\left\langle \Omega|T\psi a_{n}%
^{^{+}\epsilon}(k)|r;k_{1},...k_{n-1}\right\rangle \\
&  =i\lim_{t_{i}\rightarrow-\infty}\int_{t_{i}}d^{3}x\exp(ik_{n}\cdot
x)\overleftrightarrow{\partial_{0}}\left\langle \Omega|T\psi\epsilon^{\mu
}A_{\mu}^{T,in}(x)|r;k_{1},...k_{n-1}\right\rangle \\
&  =\frac{i}{\sqrt{Z_{3}}}\int d^{3}x\exp(ik_{n}\cdot x)(\square_{x}+\mu
^{2})\left\langle \Omega|T\psi\epsilon^{\mu}A_{\mu}(x)|r;k_{1},...k_{n-1}%
\right\rangle \\
&  =-\frac{i}{\sqrt{Z_{3}}}\int d^{3}x\exp(ik_{n}\cdot x)\left\langle
\Omega|T\psi\epsilon_{\mu}j^{\mu}|r;k_{1},..k_{n-1}\right\rangle ,
\end{align}
provided
\begin{align}
(\square_{x}+\mu^{2})T\psi A_{\mu}(x) &  =T\psi(\square_{x}+\mu^{2})A_{\mu
}(x)=T\psi(-j_{\mu}(x)+(1-\lambda)\partial_{\mu}^{x}(\partial\cdot
A(x))),\nonumber\\
\lambda(\square_{x}+\frac{\mu^{2}}{\lambda})(\partial\cdot A) &
=0,(\square_{x}+\mu^{2})\frac{\mu^{2}}{\lambda}\partial(\partial\cdot
A(x))=(\lambda-1)\partial_{\mu}(\partial\cdot A(x)).
\end{align}
where the electromagnetic current is
\begin{equation}
j^{\mu}(x)=-e\overline{\psi}(x)\gamma_{\mu}\psi(x),
\end{equation}
and $1/\lambda$ is a gauge fixing parameter and $\mu$ is a photon mass.In
deriving the spectral representation for photon propagator,the Stueckelberg
approach was used [1] and the realistic condition for transverse component of
incoming field is defined
\begin{equation}
A_{\mu}^{T,in}(x)=A_{\mu}^{in}(x)+\frac{\lambda^{2}}{\mu^{2}}\partial_{\mu
}\partial\cdot A^{in}(x),
\end{equation}
and assumed
\begin{equation}
A_{\mu}(x)_{t\rightarrow-\infty}\rightarrow\sqrt{Z_{3}}[A_{\mu}^{T,in}%
-z\frac{\lambda^{2}}{\mu^{2}}\partial_{\mu}\partial\cdot A^{in}(x)].
\end{equation}
Condition for matrix element in particular is
\begin{equation}
\left\langle 0|A_{\mu}(x)|1\right\rangle =\sqrt{Z_{3}}[\left\langle 0|A_{\mu
}^{T,in}(x)|1\right\rangle -z\frac{\lambda^{2}}{\mu^{2}}\partial_{\mu
}\left\langle 0|\partial\cdot A^{in}(x)|1\right\rangle ],
\end{equation}
where $z$ is a renormalization constant for londitudinal part[1].An
alternative way to derive the reduction formula is an imposing Lorentz
condition for the physical state%
\begin{equation}
\partial\cdot A^{(+)}|phys>=0.
\end{equation}
From the definition (9),(10) $T_{n}$ \ is seen to satisfy
Ward-Takahashi-identity:
\begin{equation}
k_{n\mu}T_{n}^{\mu}(r,k_{1},..k_{n})=eT_{n-1}(r,k_{1},..k_{n-1}),r^{2}=m^{2},
\end{equation}
provided by equal-time commutation relations%
\begin{align}
\partial_{\mu}^{x}T(\psi j_{\mu}(x)) &  =-e\psi(x),\nonumber\\
\partial_{\mu}^{x}T(\overline{\psi}j_{\mu}(x)) &  =e\overline{\psi}(x).
\end{align}
In the Bolch-Nordsieck approximation we have most singular contribution of
photons which are emitted from external lines.In perturbation theory one
photon matrix element is given
\begin{align}
T_{1} &  =\left\langle in|T(\psi_{in}(x),ie\int d^{3}x\overline{\psi}%
_{in}(y)\gamma_{\mu}\psi_{in}(y)A_{in}^{\mu}(y))|r;k\text{ }in\right\rangle
\nonumber\\
&  =ie\int d^{3}yd^{3}zS_{F}(x-y)\gamma_{\mu}\delta^{(3)}(y-z)\exp(i(k\cdot
y+r\cdot z))\epsilon^{\mu}(k,\lambda)U(r,s)\nonumber\\
&  =-ie\frac{(r+k)\cdot\gamma+m}{(r+k)^{2}-m^{2}}\gamma_{\mu}\epsilon^{\mu
}(k,\lambda)\exp(i(k+r)\cdot x)U(r,s),
\end{align}
where $U(r,s)$ is a four-component free particle spinor with positive
energy.$U(r,s)$ satisfies the relations
\begin{align}
(\gamma\cdot r-m)U(r,s) &  =0,\nonumber\\
\sum_{S}U(r,s)\overline{U}(r,s) &  =\frac{\gamma\cdot r+m}{2m}.
\end{align}
In this case the Ward-Takahashi-identity follows%
\begin{align}
k_{\mu}T_{1}^{\mu} &  =-ie\frac{1}{\gamma\cdot(r+k)-m}(\gamma\cdot
k)U(r,s)\nonumber\\
&  =-ieU(r,s)=eT_{0},
\end{align}
provided lowest-order Ward-identity%
\begin{equation}
\gamma\cdot k=(\gamma\cdot(r+k)-m)-(\gamma\cdot k-m).
\end{equation}
For general $T_{n}$ low-energy thorem determines the structure of non-singular
term in $k_{n}$ by the requirement of gauge invariance of total transion
amplitudes under the shift $\epsilon^{\mu}\rightarrow\epsilon^{\mu}%
+ck_{n}^{\mu}.$They give finite correction to the pole terms for examples in
the Bremuthstralung or Compton scattering.Detailed discussions are given in
ref[1] and non-pole terms are irerevant for the single paticle singularity in
four-dimension.Under the same assumption in three-dimension we have%
\begin{equation}
T_{n}|_{k_{n}^{2}=0}=e\frac{\gamma\cdot\epsilon}{\gamma\cdot(r+k_{n}%
)-m}T_{n-1}.
\end{equation}
From this relation the $n$-photon matrix element%
\begin{equation}
\left\langle \Omega|\psi(x)|r;k_{1},..,k_{n}\right\rangle \left\langle
r;,k_{1,}..,k_{n}|\overline{\psi}(0)|\Omega\right\rangle
\end{equation}
reduces to the products of lowest-order one photon matrix element%
\begin{equation}
T_{n}\overline{T_{n}}=%
{\displaystyle\prod\limits_{j=1}^{n}}
T_{1}(k_{j})T^{+}(k_{j})\gamma_{0}.
\end{equation}
In this case the spectral function $\rho(p)$ in (6) is given by exponentiation
of one-photon matrix element,which yields a infinite ladder approximation for
the propagator.In this way the spectral function is given in the followings
\begin{align}
\overline{\rho(x)} &  =\int\frac{md^{2}r}{(2\pi)^{2}r^{0}}\exp(ir\cdot
x)\exp(F),\\
F &  =\sum_{one\text{ photon}}\left\langle \Omega|\psi(x)|r;k\right\rangle
\left\langle r;k|\overline{\psi}(0)|\Omega\right\rangle \nonumber\\
&  =\int\frac{d^{3}k}{(2\pi)^{2}}\delta(k^{2})\theta(k_{0})\exp(ik\cdot
x)\sum_{\lambda,S}T_{1}\overline{T_{1}}.
\end{align}
To determine $F$ first we take the trace of $T_{1}\overline{T_{1}}$ for
simplicity in the infrared.%
\begin{align}
F &  =\int\frac{d^{3}k}{(2\pi)^{2}}\exp(ik\cdot x)\delta(k^{2})\theta
(k_{0})\nonumber\\
&  \times\frac{e^{2}}{4}tr\left[  \frac{(r+k)\cdot\gamma}{(r+k)^{2}-m^{2}%
}\gamma^{\mu}\frac{r\cdot\gamma+m}{2m}\gamma^{\nu}\frac{(r+k)\cdot
r}{(r+k)^{2}-m^{2}}\Pi_{\mu\nu}\right]  .
\end{align}
Here $\Pi_{\mu\nu}$ is the polarization sum
\begin{equation}
\Pi_{\mu\nu}=\sum_{\lambda}\epsilon_{\mu}(k,\lambda)\epsilon_{\nu}%
(k,\lambda)=-g_{\mu\nu}-(d-1)\frac{k_{\mu}k_{\nu}}{k^{2}},
\end{equation}
and the free photon propagator is%
\begin{equation}
D_{0}^{\mu\nu}=\frac{1}{k^{2}+i\epsilon}[g_{\mu\nu}+(d-1)\frac{k_{\mu}k_{\nu}%
}{k^{2}}].
\end{equation}
We get%
\begin{equation}
F=-e^{2}\int\frac{d^{3}k}{(2\pi)^{3}}\exp(ik\cdot x)\theta(k_{0})[\delta
(k^{2})(\frac{m^{2}}{(r\cdot k)^{2}}+\frac{1}{r\cdot k})+(d-1)\frac
{\delta(k^{2})}{k^{2}}].
\end{equation}
The second term $\delta(k^{2})/k^{2}$ equals to $-\delta^{\prime}(k^{2}).$Our
calculation is the same with the evaluation of the imaginay part of the photon
propagator.To aviod the infrared divergences which arises in the phase space
integral,we must introduce small photon mass $\mu$ as an infrared
cut-off$.$Therefore (28) is modified to%
\begin{align}
F &  =-e^{2}\int\frac{d^{3}k}{(2\pi)^{3}}\exp(ik\cdot x)\theta(k_{0}%
)\nonumber\\
&  \times\lbrack\delta(k^{2}-\mu^{2})(\frac{m^{2}}{(r\cdot k)^{2}}+\frac
{1}{(r\cdot k)})-(d-1)\frac{\partial}{\partial k^{2}}\delta(k^{2}-\mu^{2})].
\end{align}
Here we assume $\rho_{1}(\omega^{2})/m=\rho_{2}(\omega^{2})=\rho$ which is
valid in the infrared(i.e.$\gamma\cdot r=m).$In general case there are two
kinds of spectral function which is given in the appendix.It is helpful to use
function $D_{+}(x)$ to determine $F$
\begin{align}
D_{+}(x) &  =\frac{1}{(2\pi)^{2}i}\int\exp(ik\cdot x)d^{3}k\theta(k^{0}%
)\delta(k^{2}-\mu^{2})\nonumber\\
&  =\frac{1}{(2\pi)^{2}i}\int_{0}^{\infty}J_{0}(k\left\vert x\right\vert
)\frac{\pi kdk}{2\sqrt{k^{2}+\mu^{2}}}=\frac{\exp(-\mu\left\vert x\right\vert
)}{8\pi i\left\vert x\right\vert },\nonumber\\
\left\vert x\right\vert  &  =\sqrt{-x^{2}}.
\end{align}
If we use parameter trick
\begin{align}
\lim_{\epsilon\rightarrow0}\int_{0}^{\infty}d\alpha\exp(i(k+i\epsilon
)\cdot(x+\alpha r)) &  =\frac{i\exp(ik\cdot x)}{k\cdot r},\\
\lim_{\epsilon\rightarrow0}\int_{0}^{\infty}\alpha d\alpha\exp(i(k+i\epsilon
)\cdot(x+\alpha r)) &  =-\frac{\exp(ik\cdot x)}{(k\cdot r)^{2}},
\end{align}
the function $F$ is written in the following form%

\begin{align}
F  &  =ie^{2}m^{2}\int_{0}^{\infty}\alpha d\alpha D_{+}(x+\alpha r,\mu
)-e^{2}\int_{0}^{\infty}d\alpha D_{+}(x+\alpha r,\mu)-ie^{2}(d-1)\frac
{\partial}{\partial\mu^{2}}D_{+}(x,\mu)\nonumber\\
&  =\frac{e^{2}m^{2}}{8\pi r^{2}}(-\frac{\exp(-\mu\left\vert x\right\vert
)}{\mu}+\left\vert x\right\vert \operatorname{Ei}(\mu\left\vert x\right\vert
))-\frac{e^{2}}{8\pi\sqrt{r^{2}}}\operatorname{Ei}(\mu\left\vert x\right\vert
)+(d-1)\frac{e^{2}}{8\pi\mu}\exp(-\mu\left\vert x\right\vert )\\
&  =F_{1}+F_{2}+F_{g},
\end{align}
where the function $\operatorname{Ei}(\mu\left\vert x\right\vert )$ is
defined
\begin{equation}
\operatorname{Ei}(\mu\left\vert x\right\vert )=\int_{1}^{\infty}\frac
{\exp(-\mu\left\vert x\right\vert t)}{t}dt.
\end{equation}
It is understood that all terms which vanishes with $\mu\rightarrow0$ are
ignored.The leading non trivial contributions to $F$ are
\begin{equation}
\operatorname{Ei}(\mu\left\vert x\right\vert )=-\gamma-\ln(\mu\left\vert
x\right\vert )+O(\mu\left\vert x\right\vert ),
\end{equation}

\begin{align}
F_{1}  &  =\frac{e^{2}m^{2}}{8\pi r^{2}}\left(  -\frac{1}{\mu}+\left\vert
x\right\vert (1-\ln(\mu\left\vert x\right\vert )-\gamma)\right)
+O(\mu),\nonumber\\
F_{2}  &  =\frac{e^{2}}{8\pi\sqrt{r^{2}}}(\ln(\mu\left\vert x\right\vert
)+\gamma)+O(\mu),\nonumber\\
F_{g}  &  =\frac{e^{2}}{8\pi}(\frac{1}{\mu}-\left\vert x\right\vert
)(d-1)+O(\mu),
\end{align}%
\begin{equation}
F=\frac{e^{2}}{8\pi\mu}(d-2)+\frac{\gamma e^{2}}{8\pi m}+\frac{e^{2}}{8\pi
m}\ln(\mu\left\vert x\right\vert )-\frac{e^{2}}{8\pi}\left\vert x\right\vert
\ln(\mu\left\vert x\right\vert )-\frac{e^{2}}{8\pi}\left\vert x\right\vert
(d-2+\gamma),
\end{equation}
where $\gamma$ is Euler's constant and we set $r^{2}=m^{2}.$Here we used
integrals for intermediate state for on-shell fermion
\begin{align}
\int d^{3}x\exp(-ip\cdot x)\int d^{3}r\delta(r^{2}-m^{2})\exp(ir\cdot x)f(r)
&  =f(m),\\
\int d^{3}x\exp(-ip\cdot x)\int d^{3}r\frac{\exp(-mx)}{4\pi x}\exp(ir\cdot
x)\delta(r^{2}-m^{2})  &  =\frac{1}{m^{2}+p^{2}}.
\end{align}
Exponentiation of $F$ in(38) leads simple form
\begin{equation}
\exp(F)=\exp(A-B\left\vert x\right\vert )(\mu\left\vert x\right\vert
)^{D-C\left\vert x\right\vert },
\end{equation}
where\qquad\
\begin{equation}
A=\frac{e^{2}}{8\pi\mu}(d-2)+\frac{\gamma e^{2}}{8\pi m},B=\frac{e^{2}}{8\pi
}(d-2+\gamma),C=\frac{e^{2}}{8\pi},D=\frac{e^{2}}{8\pi m}.
\end{equation}
and we get the spectral function $\overline{\rho(x)}$
\begin{align}
\frac{m\exp(-m\left\vert x\right\vert )}{4\pi\left\vert x\right\vert }  &
=\int\frac{md^{2}r}{(2\pi)^{2}\sqrt{r^{2}+m^{2}}}\exp(ir\cdot x),\\
\overline{\rho(x)}  &  =\frac{m\exp(-m\left\vert x\right\vert )}%
{4\pi\left\vert x\right\vert }\exp(F(m,x))\nonumber\\
&  =\frac{m\exp(-(m_{0}+B)\left\vert x\right\vert )}{4\pi\left\vert
x\right\vert }\exp(A)(\mu\left\vert x\right\vert )^{-C\left\vert x\right\vert
+D},
\end{align}
where $m=|m_{0}+\frac{e^{2}}{8\pi}(d-2+\gamma)|,B$ denotes the correction of
mass,$D$ is a coefficent of the bare Coulomb potential devided by $m$ and
$C\left\vert x\right\vert \ln(\mu\left\vert x\right\vert )$ can be understood
as the position dependent self-energy as dynamical mass.Here $\exp
(-m\left\vert x\right\vert )/4\pi\left\vert x\right\vert $ is a free scalar
propagator with physical mass $m$.In Euclidean space we omitt the linear
infrared divergent factor $A.$

\subsection{Confining property}

\bigskip Here we mention the confining property of the propagator $S_{F}(x)$
in position space%
\begin{align}
S_{F}(x)  &  =(\frac{i\gamma\cdot\partial}{m}+1)[\frac{m\exp(-m\left\vert
x\right\vert )}{4\pi\left\vert x\right\vert }(\mu\left\vert x\right\vert
)^{D-C\left\vert x\right\vert )}]\nonumber\\
&  =(\frac{i\gamma\cdot\partial}{m}+1)\overline{\rho(x)}\\
D  &  =\frac{e^{2}}{8\pi m},C=\frac{e^{2}}{8\pi}.
\end{align}
Since $(\mu\left\vert x\right\vert )$ is a dimensionless quantity and taking
limit $\mu\rightarrow0$ with finite $\left\vert x\right\vert $ gives zero
except for $\left\vert x\right\vert =\infty$.In section III we take the
$\mu\rightarrow0$ limit in momentum space.Thus here we fix $\mu=unit$ of mass
as $m$ and see $x$ dependence of the function $(\mu\left\vert x\right\vert
)^{D-C\left\vert x\right\vert )}.$The $\overline{\rho(x)}$ damps strongly at
large $x$ provided
\begin{equation}
\lim_{x\rightarrow\infty}(\mu\left\vert x\right\vert )^{-C\left\vert
x\right\vert }=0.
\end{equation}
The profiles of the $\overline{\rho(x)}$ for various values of $D\geq1$ are
shown in Fig.1.The effect of $(\mu\left\vert x\right\vert )^{-C\left\vert
x\right\vert }$ in position space is seen to decrease the value of the
propagator at low energy and shown in Fig.2.%

{\parbox[b]{2.3289in}{\begin{center}
\fbox{\includegraphics[
height=2.3289in,
width=2.3289in
]%
{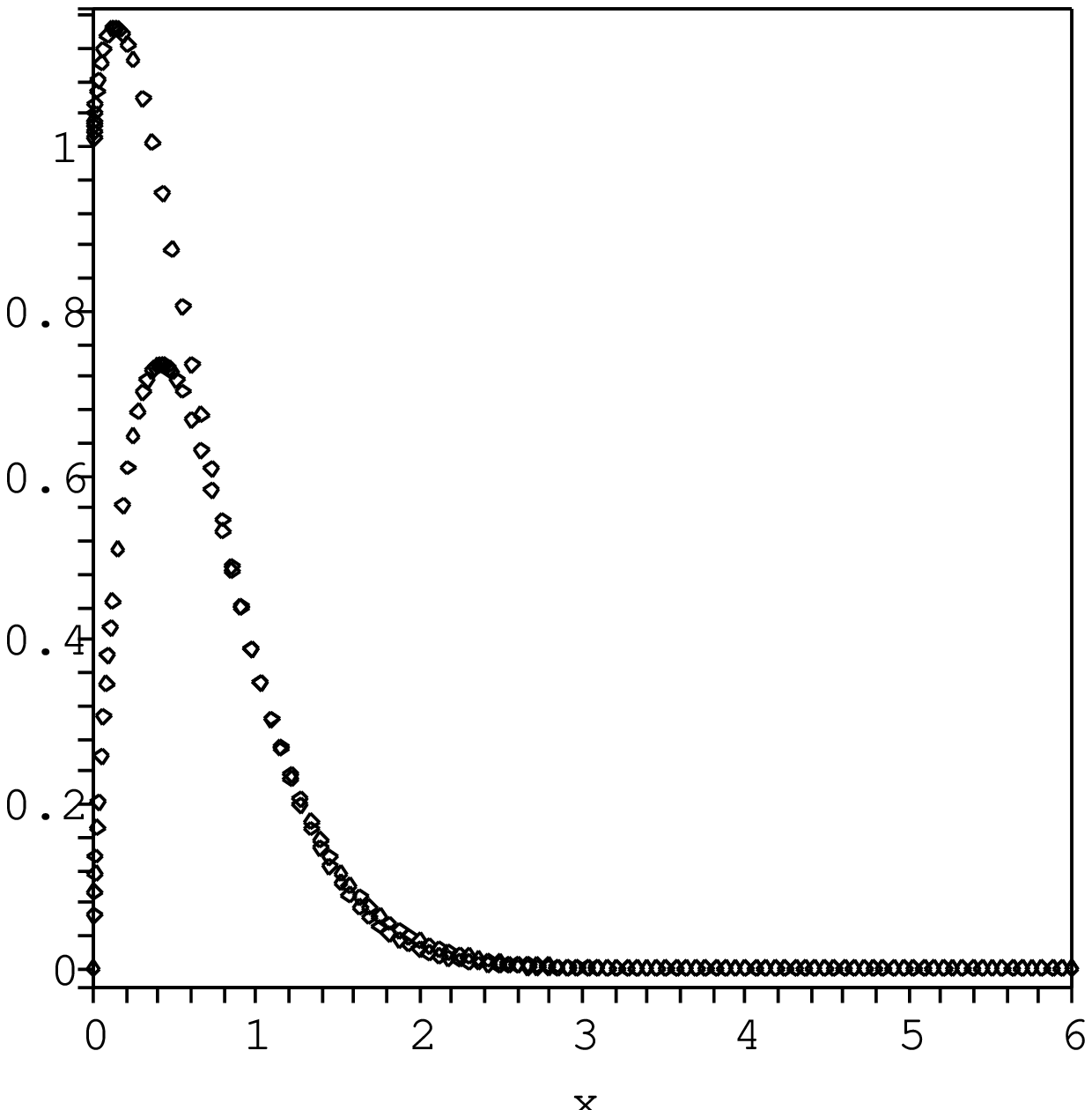}%
}\\
Fig.1 $\overline{\rho(x)}$ for $m=\mu=unit,D=1,1.5$%
\end{center}}}
{\parbox[b]{2.3272in}{\begin{center}
\fbox{\includegraphics[
height=2.3272in,
width=2.3272in
]%
{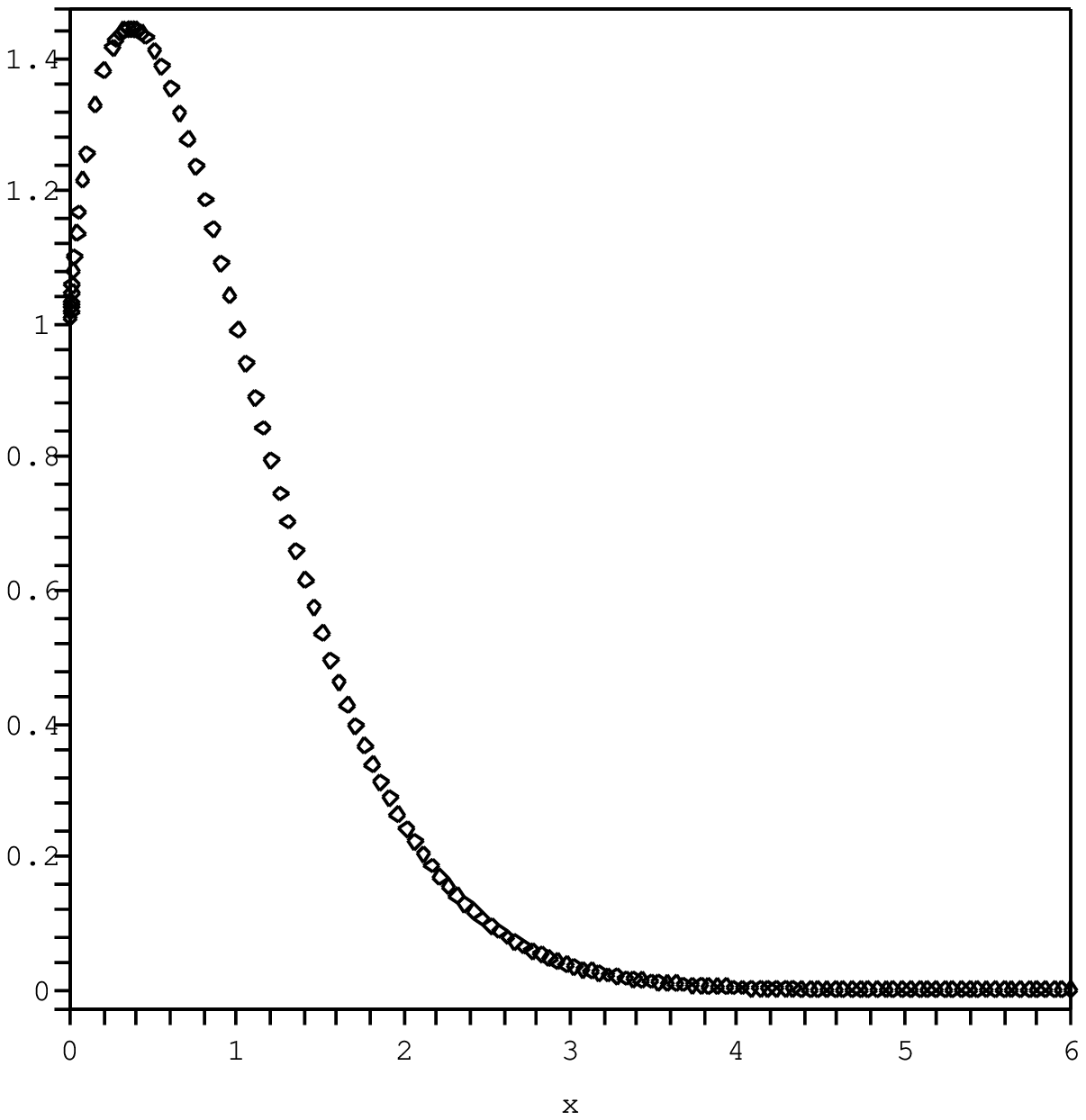}%
}\\
Fig.2 $(\mu\left\vert x\right\vert )^{-Dm\left\vert x\right\vert }$ for
$m=\mu=unit,D=1$%
\end{center}}}

In section III and IV we discuss dynamical mass,the renormalization
constant,and bare mass in connection of each terms in $F$.

\subsection{$O(e^{2})$ propagator in Momentum space}

After angular integration we get the propagator%

\begin{align}
S_{F}(p)  &  =(\frac{\gamma\cdot p}{m}+1)\rho(p),\\
\rho(p)  &  =\frac{m}{2\pi\sqrt{-p^{2}}}\int_{0}^{\infty}d\left\vert
x\right\vert \sin(\sqrt{-p^{2}}\left\vert x\right\vert )\exp(A-(m_{0}%
+B)\left\vert x\right\vert )(\mu\left\vert x\right\vert )^{-C\left\vert
x\right\vert +D},
\end{align}
where
\begin{equation}
A=\frac{e^{2}}{8\pi\mu}(d-2)+\frac{\gamma e^{2}}{8\pi m},B=\frac{e^{2}}{8\pi
}(d-2+\gamma),C=\frac{e^{2}}{8\pi},D=\frac{e^{2}}{8\pi m}.
\end{equation}
If we discuss the Euclidean or off-shell propagator we can omitt the linear
infrared divergent part in $A.$In this case $m$ denotes a physical mass
\begin{equation}
m=|m_{0}+\frac{e^{2}}{8\pi}(d-2+\gamma)|.
\end{equation}
Here we show the propagator $\rho(p)$ up to $O(e^{2})$
\begin{align}
\rho^{(2)}(p)  &  =\int d^{3}x\exp(ip\cdot x)\int\frac{md^{2}r}{r^{0}}%
\exp(ir\cdot x)F(x)\nonumber\\
&  =\frac{m}{\sqrt{-p^{2}}}\int_{0}^{\infty}d\left\vert x\right\vert
\sin(\sqrt{-p^{2}}\left\vert x\right\vert )\exp(-m\left\vert x\right\vert
)[1+A-Cx\ln(\mu\left\vert x\right\vert )+D\ln(\mu\left\vert x\right\vert
)]\nonumber\\
&  =[\frac{m(1+A)}{m^{2}+p^{2}}+m(DI_{1}-CI_{2})],
\end{align}
where $I_{1},I_{2}$ are the following integrals
\begin{align}
I_{1}  &  =\int_{0}^{\infty}\frac{\sin(\sqrt{-p^{2}}\left\vert x\right\vert
)\exp(-m\left\vert x\right\vert )}{\sqrt{-p^{2}}}\ln(\mu\left\vert
x\right\vert )d\left\vert x\right\vert \nonumber\\
&  =\frac{-\gamma}{m^{2}+p^{2}}-\frac{\ln((m^{2}+p^{2})/\mu^{2})}%
{2(m^{2}+p^{2})}-\frac{\ln((m-\sqrt{-p^{2}})/(m+\sqrt{-p^{2}}))}{m^{2}+p^{2}%
},\\
I_{2}  &  =\int_{0}^{\infty}\frac{\sin(\sqrt{-p^{2}}\left\vert x\right\vert
)\exp(-m\left\vert x\right\vert )}{\sqrt{-p^{2}}}\left\vert x\right\vert
\ln(\mu\left\vert x\right\vert )d\left\vert x\right\vert \nonumber\\
&  =\frac{-m}{(m^{2}+p^{2})^{2}}[\ln((m-\sqrt{-p^{2}})/(m+\sqrt{-p^{2}}%
))+\ln((m^{2}+p^{2})/\mu^{2})-2(1-\gamma)].
\end{align}
From these expressions we see that the gauge dependent terms $A,D,B,C,$are
wave function renormalization constant and mass renormalization
respectively.In this order the wave function renormalization constant%
\begin{equation}
Z_{2}^{(2)}=1+A-\frac{e^{2}}{8\pi m}(\gamma+\frac{1}{2}\ln((m^{2}+p^{2}%
)/\mu^{2})+\ln((m-\sqrt{-p^{2}})/(m+\sqrt{-p^{2}}))
\end{equation}
is divergent at $p^{2}=-m^{2}$ and $p^{2}=\infty.$

\subsection{\bigskip Mass generation}

Since our model $QED_{3}$ is super renormalizable mass generation occurs
perturbatively.In other words,mass changes as in the case of operator
insertion[8].It is seen by the dimensional analysis of the propagator or
self-energy $\Sigma$.For instance we take an example for the $O(e^{2})$
self-energy%
\begin{align}
\Sigma^{(2)}  &  =\gamma\cdot pA(p)+B(p),\nonumber\\
\Sigma^{(2)}  &  =e^{2}\int\frac{d^{3}k}{(2\pi)^{3}}\gamma_{\mu}S_{F}%
(k)\gamma_{\nu}D_{F0}^{\mu\nu}(p-k),\\
S_{F}(p)  &  =\frac{\gamma\cdot pA(p)+B(p)}{A^{2}(p)p^{2}+B^{2}(p)},
\end{align}
where$\Sigma^{(2)\text{ \ }}$is given by setting $A=1$,$B=m$ for $S_{F}.$By
using contour integral we obtain%
\begin{align}
\frac{1}{2p^{2}}Tr[(\gamma\cdot p)\Sigma^{(2)})  &  =A(p)=-\frac{de^{2}m}{8\pi
p^{2}}(1-\frac{p^{2}-m^{2}}{mp}\tan^{-1}(\frac{p}{m})),\\
\frac{1}{2}tr(\Sigma^{(2)})  &  =B(p)=\frac{(d-2)e^{2}m}{4\pi p}\tan
^{-1}(\frac{p}{m}).
\end{align}
Here we notice that in the limit $B(p)_{p\rightarrow0}=(d-2)e^{2}/(4\pi)$
which is independent of $m.$From the above expressions we see that the
high-energy behaviour of $A(p)$ and $B(p)$ are proportional to $e^{2}/p$ in
this order with bare vertex.On the other hand,Dyson-Schwinger equation is
non-linear%
\begin{align}
B(p)  &  =2e^{2}\int\frac{d^{3}k}{(2\pi)^{3}}\frac{1}{(p-k)^{2}}\frac
{B(k)}{k^{2}+B(k)^{2}}\nonumber\\
&  =\frac{e^{2}}{2\pi^{2}p}\int_{0}^{\infty}\frac{kdkB(k)}{k^{2}+B(k)^{2}}%
\ln(\frac{p+k}{p-k}),
\end{align}
in the Landau gauge $A=1$.This yields $B(p)\rightarrow m^{3}/p^{2}$ as
$p\rightarrow\infty.$We have the scalar part of the propagator%
\begin{equation}
S_{E}(p)=\frac{1}{4}trS_{F}(p)_{p\rightarrow\infty}\propto\frac{e^{4}}{p^{4}},
\end{equation}
which is the correct form of $B(p)$ from dimensional analysis$[8]$.Here we
return to our approximation. The propagator in momentum space is
\begin{align}
S_{F}(p)  &  =-\int d\omega^{2}\frac{\gamma\cdot p\rho_{1}(\omega^{2}%
)+\rho_{2}(\omega^{2})}{p^{2}+\omega^{2}+i\epsilon}=(\frac{\gamma\cdot p}%
{m}+1)\rho(p),\\
\rho(p)  &  =\int d^{3}x\exp(-ip\cdot x)\overline{\rho(x)},\\
\overline{\rho(x)}  &  =\frac{m\exp(-m\left\vert x\right\vert )}%
{4\pi\left\vert x\right\vert }(\mu\left\vert x\right\vert )^{-C\left\vert
x\right\vert +D}.
\end{align}
From the above equations the conditions for mass generation is easily seen
that
\begin{align}
0  &  \leq\overline{\rho(x)}\leq Max,for\text{ }1\leq D,\\
\rho(p  &  =\infty)=\lim_{p\rightarrow\infty}\frac{1}{\sqrt{-p^{2}}}\int
_{0}^{\infty}x^{2}\frac{\sin(\sqrt{-p^{2}}\left\vert x\right\vert
)}{\left\vert x\right\vert }\overline{\rho(x)}=\lim_{p\rightarrow\infty}%
\frac{1}{\sqrt{-p^{2}}}\pi(x^{2}\overline{\rho(x)})_{x=0}=0,\\
\rho(p  &  =0)=\int d^{3}x\overline{\rho(x)}=finite(\simeq0.53\text{
for}D=1,m=\mu=1),
\end{align}
which are a prioli assumed to be satisfied in the numerical analysis of
Dyson-Schwinger equation.

\section{Analysis in momentum space}

To search the infrared behaviour we expand the propagator in the coupling
constant $e^{2}$ and obtained the Fourier transform of $\overline{\rho(x)}$
[2].In that case it is not enough to see the structure of infrared behaviour
which can be compared to the well-known four dimensional QED.Instead we make
Laplace transformation of $(\mu\left\vert x\right\vert )^{-C\left\vert
x\right\vert },$which leads the general spectral representation of the
propagator in momentum space.After that we show the roles of Coulomb energy
and position dependent mass.The former determines the dimension of the
propagator and the latter acts to change mass.Let us begin to study the effect
of position dependent mass(Self-energy),Coulomb energy in momentum space%
\begin{align}
\exp(-\left\vert x\right\vert M(x))  &  =\exp(-\frac{e^{2}}{8\pi}\left\vert
x\right\vert \ln(\mu\left\vert x\right\vert ),\\
\exp(\frac{Coulomb\text{ }energy}{m})  &  =\exp(\frac{e^{2}}{8\pi m}\ln
(\mu\left\vert x\right\vert )).
\end{align}
Similar discussion was given to study the effects of self-energy and bare
potential in the stability of massless $e^{+}e^{-}$ composite in the lattice
simulation[11].The position space free propagator
\begin{equation}
S_{F}(x,m_{0})=-(i\gamma\cdot\partial+m_{0})\frac{\exp(-m_{0}\left\vert
x\right\vert )}{4\pi\left\vert x\right\vert }%
\end{equation}
is modified by these two terms which are related to dynamical mass and wave
function renormalization.To see this let us think about position space
propagator%
\begin{equation}
\overline{\rho(x)}=\frac{m\exp(-m\left\vert x\right\vert )}{4\pi\left\vert
x\right\vert }(\mu\left\vert x\right\vert )^{-\frac{e^{2}}{8\pi}\left\vert
x\right\vert }(\mu\left\vert x\right\vert )^{\frac{e^{2}}{8\pi m}}.
\end{equation}
It is easy to see that the probability of particles which are separated with
each other in the long distance is supressed by the factor $(\mu\left\vert
x\right\vert )^{-C\left\vert x\right\vert },$and the Coulomb energy modifies
the short distance behaviour from the bare $1/\left\vert x\right\vert $ to
$1/\left\vert x\right\vert ^{1-D}$.The effect of Coulomb energy for the
infrared behaviour of the free particle with mass $m$ can be seen by its
fourier transform[2,10]
\begin{align}
\int d^{3}x\exp(-ip\cdot x)\frac{\exp(-m\left\vert x\right\vert )}%
{4\pi\left\vert x\right\vert }(\mu\left\vert x\right\vert )^{D}  &  =\mu
^{D}\frac{\Gamma(D+1)\sin((D+1)\arctan(\sqrt{-p^{2}}/m))}{\sqrt{-p^{2}}%
(p^{2}+m^{2})^{(1+D)/2}}\nonumber\\
&  \sim\mu^{D}(\sqrt{-p^{2}}-m)^{-1-D}\text{ near }p^{2}=-m^{2}.
\end{align}
Above formula shows the structure in momentum space is modified for both
infrared and ultraviolet regions.Usually constant $D$ represents the
coefficent of the leading infrared divergence for fixed mass in four
dimension.Therefore Coulomb energy has the same effects in three dimension as
in four dimension but change the ultraviolet behaviour since the coupling
constant $e^{2}$ is not renormalized.Now we consider the role of $M(x)$ as the
dynamical mass at low momentum.First we define Fourier transform of the
$\overline{\rho(x)}$;
\begin{equation}
\rho(p)=\int\exp(-ip\cdot x)\overline{\rho(x)}d^{3}x.
\end{equation}
Momentum dependence of $\rho(p)$ for various values of $D$ is shown
numerically in Fig.3.We notice that
\begin{align}
D  &  =0\rightarrow\rho(p)=\frac{m}{p^{2}+m^{2}},\\
D  &  =1\rightarrow\rho(p)=\frac{2m^{2}\mu}{(p^{2}+m^{2})^{2}}\times\log\text{
correction,}%
\end{align}
which we can see from (70),(58)-(60).If we include $(\mu\left\vert
x\right\vert )^{-C\left\vert x\right\vert }$ term it is easy to see that the
value of the proagator $\rho(p=0)$ decreases which is shown numerically in
Fig.3.In the previous paper we expand the propagator $\rho(p)$ in powers of
$e^{2}$ and see the logarithmic infrared divergence at $p^{2}=m^{2}$.In that
case it is not clear the effect of position dependent mass.If we use Laplace
transformation it is easily seen that $(\mu\left\vert x\right\vert
)^{-C\left\vert x\right\vert }$ acts as mass changing operator $m\rightarrow
m-s$
\begin{align}
F(s)  &  =\int_{0}^{\infty}\exp(-s\left\vert x\right\vert )(m\left\vert
x\right\vert )^{-C\left\vert x\right\vert }d\left\vert x\right\vert
,\nonumber\\
(\exp(-m^{\ast}\left\vert x\right\vert )(m\left\vert x\right\vert
)^{-C\left\vert x\right\vert }  &  =\int_{0}^{\infty}\exp(-(m^{\ast
}-s)\left\vert x\right\vert )F(s)ds,
\end{align}
where we introduced%
\begin{equation}
m^{\ast}=m+C\ln(\frac{\mu}{m}),
\end{equation}
to separate the cut-off $\mu$ dependence from $F(s)$ in the following way
\begin{equation}
\exp(-m\left\vert x)\right\vert (\mu\left\vert x\right\vert ^{-C\left\vert
x\right\vert }=\exp(-m^{\ast}\left\vert x\right\vert )(m\left\vert
x\right\vert )^{-C\left\vert x\right\vert }.
\end{equation}
The Laplace transfom $F(s)$ is shown in Fig.4.%

{\parbox[b]{2.4958in}{\begin{center}
\fbox{\includegraphics[
height=2.4958in,
width=2.4958in
]%
{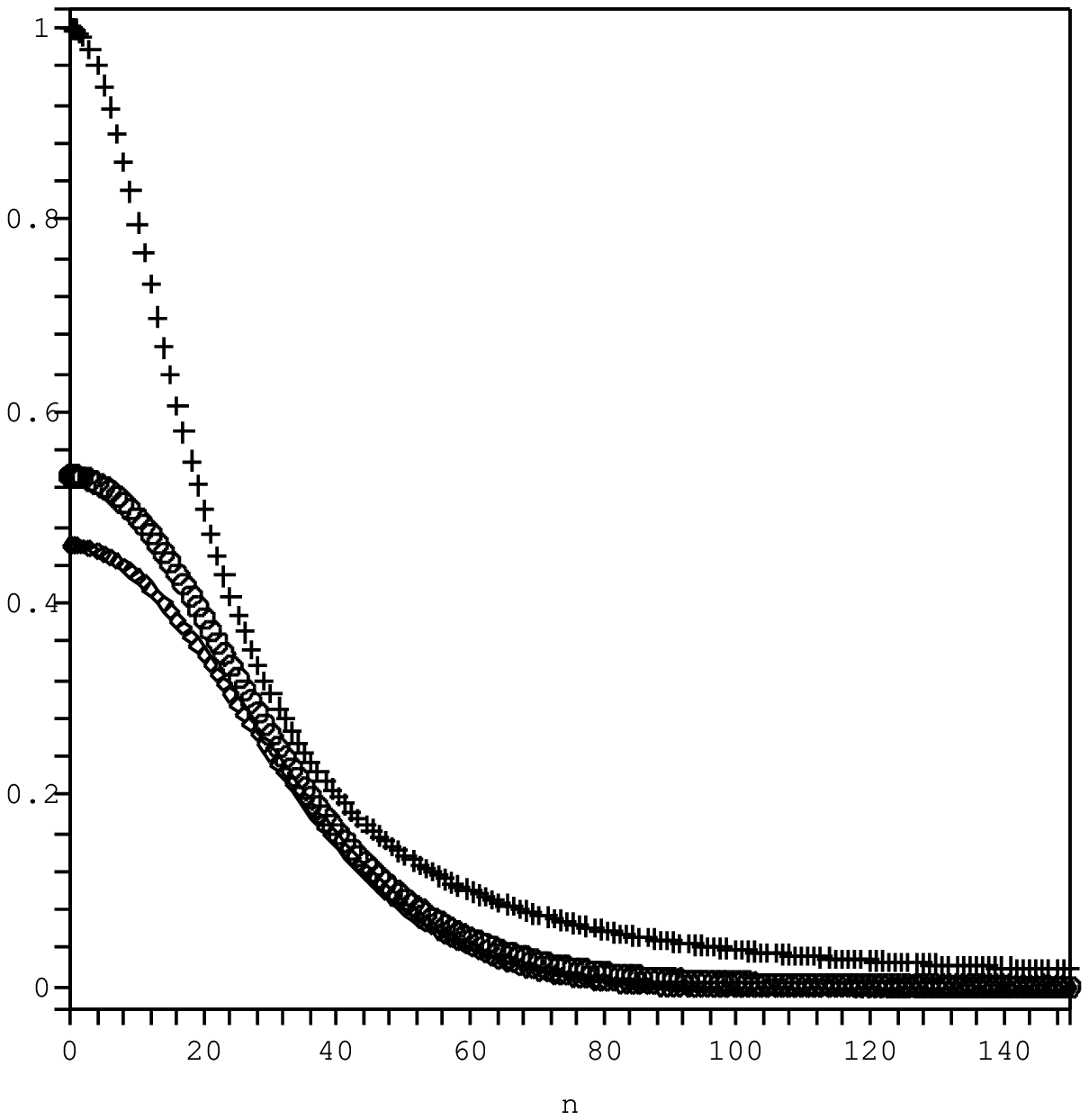}%
}\\
Fig.3 $\rho(p)$ for $m=\mu=unit,D=0(upper),1(middle),1.5(upper),p=n/20$%
\end{center}}}
{\parbox[b]{2.4958in}{\begin{center}
\fbox{\includegraphics[
height=2.4958in,
width=2.4958in
]%
{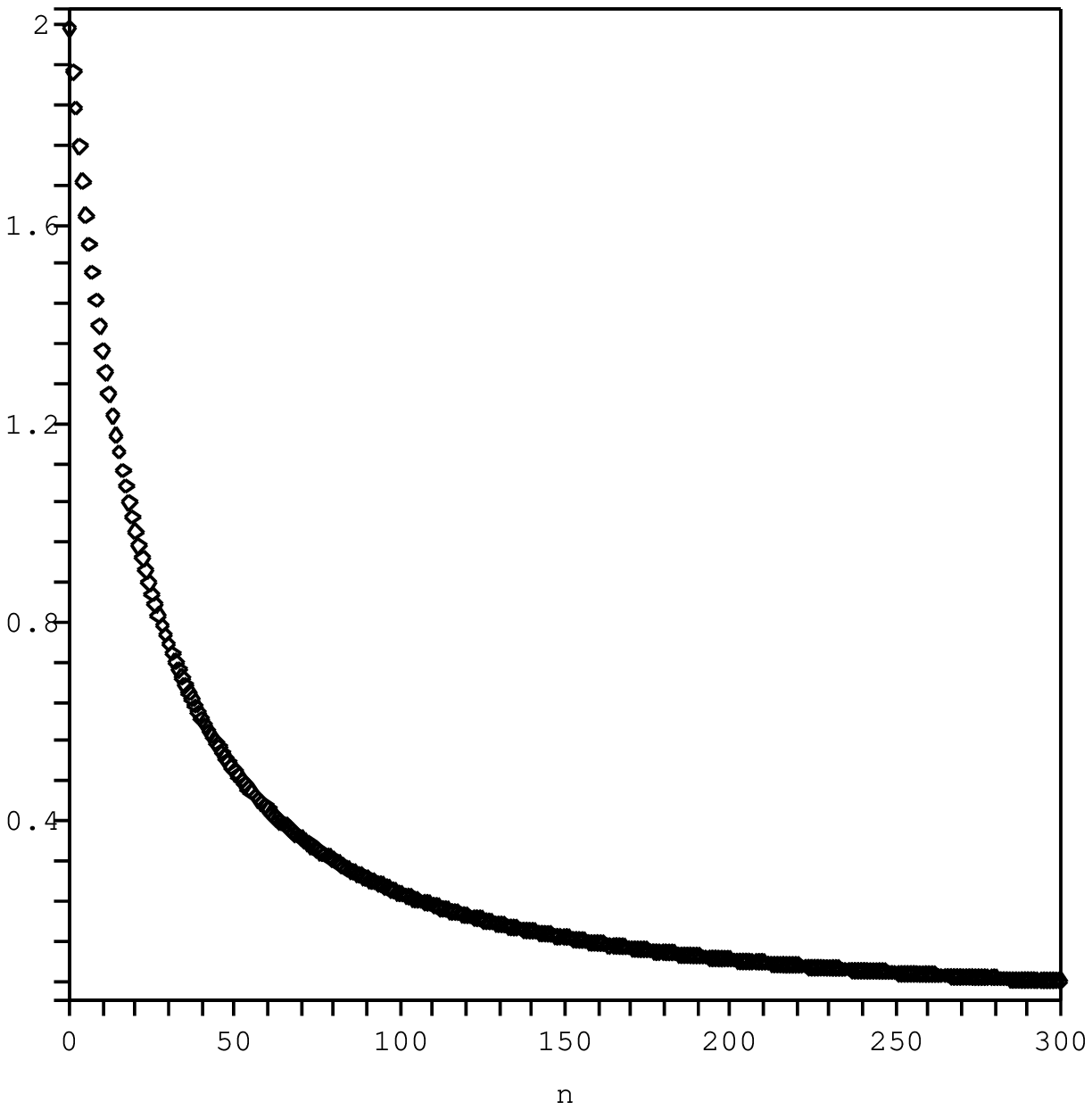}%
}\\
Fig.4 Laplace transform $F(s)$ for $(m\left\vert x\right\vert )^{-Dm\left\vert
x\right\vert },$ $m=1,D=1,s=n/20$%
\end{center}}}

\bigskip

We have the complete the expression of the propagator%
\begin{equation}
\overline{\rho(x)}=\frac{m\mu^{D}\exp(-m^{\ast}\left\vert x\right\vert )}%
{4\pi\left\vert x\right\vert ^{1-D}}\int_{0}^{\infty}\exp(s\left\vert
x\right\vert )F(s)ds,
\end{equation}%
\begin{align}
\rho(p)  &  =m\mu^{D}\int_{0}^{\infty}\frac{\Gamma(D+1)\sin((D+1)\arctan
(\sqrt{-p^{2}}/(m^{\ast}-s)))F(s)ds}{\sqrt{-p^{2}}\sqrt{(p^{2}+(s-m^{\ast
})^{2})^{1+D}}},\\
\rho(p  &  =0)=\int_{0}^{\infty}\frac{2(s-m^{\ast})F(s)ds}{(s-m^{\ast})^{4}%
}=\frac{1}{2\pi i}\int_{s-i\infty}^{s+i\infty}\frac{2F(s)ds}{(s-m^{\ast})^{3}%
}\simeq0.53\text{ for }m=\mu=1,D=1.
\end{align}
Here we study the cut off $\mu\rightarrow0$ limit. We see that $m^{\ast}$
diverges as $\ln(\mu/m)$ in the limit $\mu\rightarrow0.$ In this case we can
neglect the $s$ dependence of the denominator of $\rho(p)$ and using%
\begin{align}
0  &  \leq\frac{s-m^{\ast}}{((p^{2}+(s-m^{\ast})^{2})^{\frac{1+D}{2}}}%
\leq\frac{s-m^{\ast}}{(p^{2}+m^{\ast2})^{\frac{!+D}{2}}},\nonumber\\
\int_{0}^{\infty}F(s)ds  &  =1,\int_{0}^{\infty}sF(s)ds=finite,
\end{align}
we get the infrared behaviour%
\begin{equation}
\rho(p)\thickapprox\frac{m^{1+D}\epsilon^{D}\Gamma(1+D)\sin((D+1)\arctan
(\sqrt{-p^{2}}/(s-m^{\ast}))}{\sqrt{-p^{2}}(p^{2}+(s-m^{\ast})^{2}%
)^{\frac{1+D}{2}}}\rightarrow0,\epsilon=\frac{\mu}{m}\rightarrow0,
\end{equation}
and for $D=1$%
\begin{equation}
\rho(p)\thickapprox\frac{m^{2}\epsilon(1+\ln(\epsilon))}{(p^{2}+m^{2}%
(1+\ln(\epsilon))^{2})^{2}}\rightarrow0.
\end{equation}
In Minkowski space we have by replacement $p^{2}\rightarrow-p^{2}$ with
infrared factor $\exp(A),$%
\begin{equation}
A=\frac{e^{2}}{8\pi\mu}(d-2)+\frac{e^{2}}{8\pi m}.
\end{equation}
Since $A$ is highly gauge dependent we do not have a definite answer in the
limit $\mu\rightarrow0$ except for the Yennie gauge $d=2$.The infrared
behaviour near $p^{2}=m^{2}$ becomes
\[
\rho(p)\thickapprox\frac{\Gamma(D+1)\epsilon^{D}\sin((D+1)\arctan
(p/m(1+D\ln(\epsilon)))}{(-p^{2}+m^{2}(1+D\ln(\epsilon))^{2})^{1+D/2}%
}\rightarrow0,
\]
and for $D=1$%
\begin{equation}
\rho(p)\thickapprox\frac{2m^{3}\epsilon(1+\ln(\epsilon))}{(-p^{2}+m^{2}%
(1+\ln(\epsilon))^{2})^{2}}\rightarrow0.
\end{equation}
this leads to
\begin{equation}
\rho(p)\thickapprox\frac{2\epsilon}{m(\ln(\epsilon))^{4}}=\frac{2\sqrt
{1-p^{2}/m^{2}}}{m(\ln(\sqrt{1-p^{2}/m^{2}}))^{4}},\text{ for }D=1,d=2,
\end{equation}
near $p^{2}=m^{2}.$In Euclidean space it is natural that in the long-distance
the propagator vanishes in the limit $\mu\rightarrow0.$In Minkowski space we
find a vanishiment of the propagator as $\sqrt{1-p^{2}/m^{2}}(\ln
(1-p^{2}/m^{2}))^{-4}$ near $p^{2}=m^{2}$ for the Yennie gauge at $D=1$,which
is a consequence of confinement picture in our model.In comparison between our
model and Dyson-Schwinger equation,$D=1$ is prefered for dynamical mass
generation.It has been discussed that the propagator will have a branch point
on the real $p^{2}$ axis associated with dynamical mass[12,13].Therefore we
can say that the radiatively corrected fermion has not a simple structure as
in QED$_{4},$in QED$_{3}$ it has a superposition of different mass with cut
and linear divergence associated with massless photon in the ordinary phase or
it is expressed as superposition of dipoles with linear divergence in the
condensed phase for finite cut-off $\mu$,which will be shown in the next
section.There is a possibility to remove infrared cut-off by including photon
mass as Chern-Simon term or vacuum polarization of photon[3,5,7].

\section{ Bare mass and vacuum expectation value $\left\langle \overline{\psi
}\psi\right\rangle $}

In this section we examine the renormalization constant and bare mass and
study the condition of vanishing bare mass based on spectral
representation.The equation for the renormalization constant in terms of the
spectral functions read%

\begin{equation}
\lim_{p\rightarrow\infty}\frac{Z_{2}^{-1}(\gamma\cdot p+m_{0})}{p^{2}%
-m_{0}^{2}+i\epsilon}=\lim_{p\rightarrow\infty}\int\frac{\gamma\cdot p\rho
_{1}(\omega^{2})+\rho_{2}(\omega^{2})d\omega^{2}}{p^{2}-\omega^{2}+i\epsilon}.
\end{equation}
Instead we determine them directly by taking the high energy limit of
$S_{F}(p)$%
\begin{align}
m_{0}Z_{2}^{-1}  &  =m\int\rho_{2}(\omega^{2})d\omega^{2}=\frac{1}{4}%
\lim_{p\rightarrow\infty}tr[p^{2}S_{F}(p)]\\
Z_{2}^{-1}  &  =\int\rho_{1}(\omega^{2})d\omega^{2}=\frac{1}{4}\lim
_{p\rightarrow\infty}tr[\gamma\cdot pS_{F}(p)].
\end{align}
In (86) the function $\sin((D+1)\arctan(\sqrt{-p^{2}}/(m^{\ast}-s))$ is
expanded for large $(-$ $p^{2}),$
\begin{equation}
\sin((D+1)\arctan(\frac{\sqrt{-p^{2}}}{m^{\ast}-s})=\sin(\frac{(D+1)}{2}%
\pi)-\cos(\frac{(D+1)}{2}\pi)\frac{(D+1)(m^{\ast}-s)}{\sqrt{-p^{2}}}%
+O(\frac{1}{-p^{2}}).
\end{equation}
From this we obtain for $D<1$
\begin{align}
m_{0}Z_{2}^{-1}  &  =m\mu^{D}\Gamma(D+1)\sin(\frac{(D+1)\pi}{2})\lim
_{p^{2}\rightarrow\infty}\sqrt{-p^{2}}^{-D}=\left[
\begin{array}
[c]{cc}%
0 & (0<D)\\
m & (0=D)
\end{array}
\right]  ,\\
Z_{2}^{-1}  &  =\mu^{D}\Gamma(D+1)\sin(\frac{(D+1)\pi}{2})\lim_{p^{2}%
\rightarrow\infty}\sqrt{-p^{2}}^{-D}=\left[
\begin{array}
[c]{cc}%
0 & (0<D)\\
1 & (0=D)
\end{array}
\right]  .
\end{align}
For $D=1$ case first term in the $1/\sqrt{-p^{2}}$ expansion vanishes and
\begin{align}
m_{0}Z_{2}^{-1}  &  =\lim_{p\rightarrow\infty}\frac{2m\mu}{-p^{2}}\int
(m^{\ast}-s)F(s)ds=0,\\
Z_{2}^{-1}  &  =\lim_{p\rightarrow\infty}\frac{2\mu}{-p^{2}}\int(m^{\ast
}-s)F(s)ds=0.
\end{align}
This means that propagator in the high energy limit has no part which is
proportional to the free one and it shows confinement.Usually mass is a
parameter which appears in the Lagrangean.For example chiral symmetry is
defined for the bare quantity.In ref[9] the relation between bare mass and
renormalized mass of the fermion propagator in QED is discussed based on
renormalization group equation with the assumption of ultraviolet stagnant
point and shown that the bare mass vanishes in the high energy limit even if
we start from the finite bare mass in the theory.It suggests that symmetry
properties can be discussed in terms of renormalized quantities.In QCD bare
mass vanishes in the short distance by asymptotic freedom.And the dynamical
mass vanishes too[8].Since our model is super-reomalizable bare mass $m_{0}$
does not vanish.In our apprximation this problem is understood that at short
distance propagator in position space tends to
\begin{align}
\overline{\rho(x)}  &  =\frac{m\exp(-m_{0}\left\vert x\right\vert )}%
{4\pi\left\vert x\right\vert }\exp(-B\left\vert x\right\vert +D\ln
(\mu\left\vert x\right\vert )-C\left\vert x\right\vert \ln(\mu\left\vert
x\right\vert ))_{x\rightarrow0}\\
&  \rightarrow\left\vert x\right\vert ^{D-1}(\mu\left\vert x\right\vert
)^{-C\left\vert x\right\vert },
\end{align}
where we have $\overline{\rho(0)}=finite$ at $D=1$ case which is independent
of the bare mass $m_{0}$.Thus we have a same effect as vanishing bare mass in
four dimensional model.Of course we have a dynamical mass generation which is
$m=|\frac{e^{2}}{8\pi}(d-2+\gamma)|$ $+$ $M(x)$ for $m_{0}=0$ in our
approximation$.$There is a chiral symmetry at short distance where the bare or
dynamical mass vanishes in momentum space but its breaking must be discussed
in terms of the values of the order parameter.Therefore it is interesting to
study the possibility of pair condensation in our approximation.The vacuum
expectation value of pair condensate is evaluated
\begin{align}
\left\langle \overline{\psi}\psi\right\rangle  &  =-trS_{F}(x)=-2m\mu^{D}%
\lim_{x\rightarrow0_{+}}\frac{\exp(-m\left\vert x\right\vert )}{\left\vert
x\right\vert }(\mu\left\vert x\right\vert )^{-C\left\vert x\right\vert
+D}\nonumber\\
&  =\left(
\begin{array}
[c]{ll}%
0 & (D>1)\\
finite & (D=1)\\
\infty & (D<1)
\end{array}
\right)  ,
\end{align}
for finite cut-off $\mu.$For $D<1$ case the $\overline{\rho(x)}$ is divergent
at $x=0$ and it looks like a spike,for $D=1,\overline{\rho(0)}=finite$ and it
looks like a wave packet at finite range where long range correlation
appears,and for $D>1,\overline{\rho(0)}=0$ but it does not damp fast for large
$x$.In the weak coupling limit we obtain $Z_{2}=1,m_{0}=m$ and $\left\langle
\overline{\psi}\psi\right\rangle =\infty$.If we introduce chiral symmetry as a
global $U(2n)$,it breaks dynamically into $SU(n)\times SU(n)\times U(1)\times
U(1)$ as in QCD[8,11] for $D=1$ for finite infrared cut-off.Our model may be
applicable to relativistic model of super fluidity in three dimension.Usually
we do not find the critical coupling $D=1$ in the analysis of the
Dyson-Schwinger equation in momentum space where only continuum contributions
are taken into account and we do not define physical mass.

\section{Summary}

Infrared behaviour of the propagator has been examined in the
Bolch-Nordsieck-like approximation to QED$_{3}$.Using LSZ reduction formula
for the one photon matrix element we find two kinds of logarithmic infrared
divergent terms.These are position dependent mass and Coulomb energy both of
which are gauge independent in our approximation. Confining property is shown
by position dependent mass at large distance where propagator vanishes faster
than the sloution of Dyson-Schwinger equation converted into position space
with some approximation.Renormalization constant $Z_{2}^{-1}$ vanishes for
arbitrary coupling $D$ in our approximation,which implies confinement.Momentum
space propagator is expressed as dispersion integral and it shows us that the
effects of dynamical mass can be written as a superposition of the propagator
with different masses.In our approximation there seems to be a critical
coupling constant for vacuum expectation value $\left\langle \overline{\psi
}\psi\right\rangle $ which is independent of the bare mass.Our model shows
that the confining property of charged fermion and dynamical mass generation
are realized in $QED_{3}$ (at critical coupling constant $D=1$) as in $QCD$ in
four dimension with finite infrared cut-off which may be supplied by some
dynamical origin of the photon mass beyond the quenched approximation .

\section{\bigskip Acknowledgement}

The author would like thank Prof.Roman Jackiw at MIT and Prof.Robert Delbourgo
at University of Tasmania for their hospitalities and communications on our
subjects on infrared problems in three dimensional gauge theory.He also thanks
to Dr.Suzuki at Hokkaido University for his interest for the infrared limit.

\section{A. gauge dependence}

In ref[2,5,9] gauge dependece of the propagator has been discussed.In order to
understand the gauge dependence of the propagator away from the threshold,we
will once examine the gauge covariance relation.Writing $D_{\mu\nu}%
(k)=D_{\mu\nu}^{(0)}(k)+k_{\mu}k_{\nu}M(k),$we see that the fermion proagator
varies with the gauge according to[4],%
\begin{align}
S_{F}(x,K)  &  =\exp(-iK\sqrt{x^{2}})S_{F}(x,0),\\
M(x)  &  =ie^{2}d\int\frac{d^{3}k}{(2\pi)^{3}}\frac{1}{k^{4}}(\exp(-ik\cdot
x)-1)=-iK\sqrt{x^{2}},
\end{align}
where $K=e^{2}d/8\pi.$Thus the gauge transformation property is violated by
linear infrared divergences near the threshold.Here we adopt the subtracion of
the point $k=0$ in the integral$.$Free propagator with mass $\omega$ equals
to
\begin{equation}
S_{F}^{(0)}(x)=-(i\gamma\cdot\partial+\omega)\frac{\exp(-\omega\sqrt{-x^{2}}%
)}{4\pi\sqrt{-x^{2}}},
\end{equation}
and its Fourier transform is%
\begin{equation}
S_{F}^{(0)}(p)=\int d^{3}x\exp(ip\cdot x)S_{F}^{(0)}(x)=\frac{\gamma\cdot
p+\omega}{p^{2}-\omega^{2}}.
\end{equation}
In our approximation only mass terms changes under the gauge transformation
and remaining terms in $F$ are gauge invariant.In general gauge invariant part
is approximation dependent.

Finally we show the general spin dependent spectral function in the $O(e^{2})$
for $d$ gauge by evaluating%
\[
F_{v}=\frac{1}{4r^{2}}tr(\gamma\cdot rF),F_{s}=\frac{1}{4}tr(F),
\]
and we have%
\begin{align}
F  &  =(\frac{\gamma\cdot r}{m}+1)[\frac{e^{2}}{8\pi}(-(d-2+\gamma)\left\vert
x\right\vert -\frac{e^{2}}{8\pi}\left\vert x\right\vert \ln(\mu\left\vert
x\right\vert )+\frac{e^{2}}{8\pi m}(\ln(\mu\left\vert x\right\vert
)+\gamma)]\nonumber\\
&  -\frac{\gamma\cdot r}{m}\frac{e^{2}(d-1)}{8\pi m^{2}\left\vert x\right\vert
}.
\end{align}
The first term is the same as the previous one.The second term is not
significant in the infrared.This term is also obtained by gauge transformation
from the Landau gauge where $\rho_{1}=\rho_{2}$ is assumed[7].Since%
\[
P=\frac{\gamma\cdot r+m}{2m}%
\]
is a projection operator it is easy to show the following.In Euclidean space
if we exponentiate we have
\begin{align}
&  S_{F}(x)=-\int\frac{mrdr}{\sqrt{r^{2}+m^{2}}}\exp(ir\cdot x)(1+\frac
{r\cdot\gamma}{m})\exp((\frac{r\cdot\gamma}{m}+1)F)\nonumber\\
&  =-\int\frac{mrdr}{\sqrt{r^{2}+m^{2}}}\exp(ir\cdot x)(1+\frac{r\cdot\gamma
}{m})\exp(\frac{F}{2})(\cosh(\frac{F}{2})+\frac{r\cdot\gamma}{m}\sinh(\frac
{F}{2}))\nonumber\\
&  =-\int\frac{mrdr}{\sqrt{r^{2}+m^{2}}}\exp(ir\cdot x)(1+\frac{r\cdot\gamma
}{m})\exp(F),\\
S_{F}(p)  &  =\int d^{3}x\exp(-ip\cdot x)S_{F}(x).
\end{align}

\section{References}

\noindent\lbrack1]R.Jackiw,L.Soloviev,Phys.Rev.\textbf{173}.(1968)1458;

F.E.Low,Phys.Rev\textbf{110(}1958)974;

M.Gell-Mann,M.L.Goldberger.Phys.Rev.\textbf{96(}1954)1433;

S.Weinberg,Phys.Rev\textbf{.140(}1965)B516;

R.Jackiw,Phys.Rev.\textbf{168}(1968)1623;

L.D.Landau and E.M.Lifzhits,Quantum Electrodynamics,Pergamon \ Press,Oxford(1982);

C.Itzykson,J-B Zuber,Quantum Field Theory,McGRAW-HILL.\newline%
[2]Y.Hoshino,\textbf{JHEP05}(2003)075.\newline%
[3]S.Deser,R,Jackiw,S.Templton,Ann.Phys.(NY)\textbf{140}(1982)372.\newline%
[4]L.D.Landau,Khalatonikov,Zh.Eksp.Theor.Fiz.\textbf{29}%
(1956)89[Sov.Phys.JETP\textbf{2}(1956)69;\newline%
\ \ \ \ B.Zumino,J.Math.Phys.\textbf{1}(1960)1.\newline%
[5]Conrand.J.Burden,Justin.Praschika,Craig.D.Roberts,Phys.Rev.\textbf{D46}%
(1992)2695; \newline%
\ \ \ \ Conrad.J.Burden,Craig.D.Roberts,Phys.Rev.\textbf{D47(}1993\textbf{)}5581;

A.Bashir,A.Raya,Phys.Rev.\textbf{D64}(2001)105001.\textbf{\newline%
}[6]T.Appelquist,R.Pisarski,Phys.Rev.\textbf{D23}%
(1981)2305;T.Appelquist,U.Heinz,Phys.Rev.\textbf{D24},

(1981)2305.\newline[7]A.B.Waites,R.Delbourgo,Int.J.Mod.Physics.\textbf{A}%
.\textbf{7}(1992)6857;

E.Abdalla,R.Banerjee,C.Molina,Eur.Phys.J.C.\textbf{17}(1998)467.\newline%
[8]H.D.Politzer,Nucl.Phys.\textbf{B117}(1976)397.\newline%
[9]K.Nishijima,Prog.Theor.Phys.\textbf{81}(1989)878;\textbf{83}%
(1990)1200.\newline[10]L.S.Brown,Quantum Field Theory,Cambridege University Press(1992);

N.N.Bogoliubov,D.V.Shirkov,Introduction to the theory of Quantized

Fields,section32,Wiley-Interscience;

L.D.Landau,and Lifzhits,Quantum Electrodynamics,Pergamon
Press,Oxford(1982).\newline%
[11]T.Appelquist,D.Nash,L.C.R.Wijewardhana,Phys.Rev.Lett.\textbf{60}(1988)1338;

E.Dagotto,J.B.Kogut,A.Kocic,Phys.Rev.Lett.\textbf{62}(1988)1083;

Y.Hoshino,T.Matsuyama,Phys.Lett.\textbf{B222}(1989)493.\newline%
[12]D.Atkinson,D.W.E.Blatt,Nucl.Phys.\textbf{151B}(1979)342.\newline%
[13]M.Koopman,Dynamical Mass Generation in QED$_{3},$Ph.D thesis,Groningen
University (1990),Chapter4;P.Maris,Phys.Rev.\textbf{D52}%
(1995);Y.Hoshino,Il.Nouvo.Cim\textbf{.112A}(1999)335.\newline

\end{document}